# Architecture and Production Readiness Reviews in Practice

James Cusick
Wolters Kluwer, Corporate Legal Services
New York, NY
j.cusick@computer.org

## ABSTRACT

*In order to succeed in building and deploying complex software solutions, an architecture is essential. For many in the industry structured reviews of these architectures is also de rigor. Practices for such reviews have been developed and reported on for years. One aspect that does not receive as much attention but is no less important is the relationship between these architectures and the requirements for deploying them into production environments. At Wolters Kluwer's Corporate Legal Services we first established a typical architecture review process and then established a two phase production preparation review process. This paper describes in detail how these practices work and some of the technical results of these reviews including the frequency and style of the reviews, the process automation around them, and the number and nature of some of the technical flaws eliminated by enforcing these reviews. This paper lays the ground work for others who would be interested in following similar practices.*

## Introduction

Architecture Reviews have a long history of providing value in the software field [1]. My own experience with architecture reviews started on one of my first major projects working for the then AT&T Bell Laboratories. Our project was complex and running into multiple issues. Management called in an independent review team from the central labs organization that provided extensive analysis quickly on what the issues were and provided recommendations on how to address them.

Later while working in the Operations Technology Center of AT&T's Network Services Divsion, which had over 1,500 technical staff, I was a part of their internal review team. I learned how to plan and conduct these reviews. Still later while working for the Advanced Technologies group within the Labs I actually was responsible for running process reviews across the company and also ran some architecture reviews as well.

From all of these experiences I learned the value of independent expert reviews. In my current role I established a very light weight architecture review. I also established two production readiness reviews of which I believe the first is novel. These reviews cover production readiness and are the focus of this paper in addition to a description of our architecture review approach. I will describe the various reviews we have in place, the types of questions we ask, and the results of these reviews. From this it will be possible for others to emulate these same kinds of production readiness reviews.

# Background

The use of software architecture reviews has been well known for many years. As mentioned, at AT&T this practice was well established early on. One of the first formal publications of a structured practice was from AT&T [1]. This "blue book" was part of a series of technical best practices developed within the Labs and used throughout the company. This book reduced the concepts of the architecture review to a practice that could be followed. It focused on independent reviewers and quick analysis turn around.

In following years other companies shared their practices and the approach to software architecture review became commonly understood. Eventually the Software Engineering Institute also published a best practice in this area [2]. The common knowledge around reviews has continued to be relied on. Recently new authors have also discussed essentially these same practices [3].

In a recent detailed survey of the field of architecture reviews [4] such reviews were used in both large and small organizations, and 56% reported informal review practices. The techniques applied included experience based reasoning, prototyping, scenarios, and checklists (40%). Only 5% of respondents used models. This study also documented participant's views of the benefits of these reviews which included identifying risks, assessing quality attributes, identifying reuse opportunities, and promoting good architecture practice.

What is generally missing from these process descriptions and methods is a focus on production readiness. This is what we have developed to assist our development preparedness and we will now discuss the linkage between a fairly standard architecture review practice with a two-phase production readiness review practice. There are similar production readiness reviews in existence [5][6]. These are in the military contracting world primarily and serve the purpose of reviewing the acceptance of a solution. The concept is generally known in project management as well and especially engineering. However, I have not seen it used often or widely in software development and not in relation to an architecture review. Our approach is similar to these reviews in principle but our reviews are conducted within a commercial setting and they are used internally from development to production teams. Also, the early first pass review which we employ does not appear to be used elsewhere.

# Review Process Summary

Within Wolters Kluwer Corporate Legal Services we have established a review process flow as depicted in Figure 1 below. As can be seen, once a project is initiated, among other steps, a technical approach document is created. This document is typically the container document for the architecture. There may be additional supporting documents but this approach document will contain the intent of the system or the solution and its implementation outline. Once that document is ready for review, typically at the end of the analysis phase or after some iterations when following a Scrum methodology, a review is scheduled. Reviewers are drawn from other projects and from specialty areas like systems and database and potentially some senior technical mangers. The reviews are kept short and the goal is to validate the approach and to look for means of leveraging existing solutions or technologies. The architect is given ideas on how to improve the solution.

Following this technical/architecture review a Production Readiness Exploration (PRE) is held. This review is a non-binding exploration of the readiness of the project to migrate to production. Such issues as infrastructure needs, networking preparedness, firewall needs, storage, and the like are discussed. The concept is to prepare the team for the work they will need to do to be able to move into production beyond simply writing code. The reason it is non-binding is that the project still needs to go through design and implementation and a number of things may change. It is after that when a binding review is held.

Once construction and testing are complete and testing has started a Production Readiness Review (PRR) is held. This review is a binding review and without a pass the project cannot move into production. The review questions are similar to those in the PRE but now they need to be satisfied. If they are not then adjustments need to me made. This could be to develop a capacity model or to define the growth in storage usage expected or similar questions. Once the team has satisfied the reviewer's questions the PRR approval is added to the list of Release Management requirements satisfied to be approved for migration to production. The remainder of this paper will expand on the details of these reviews, how they are held, some of the checklists used, and the results of the reviews.

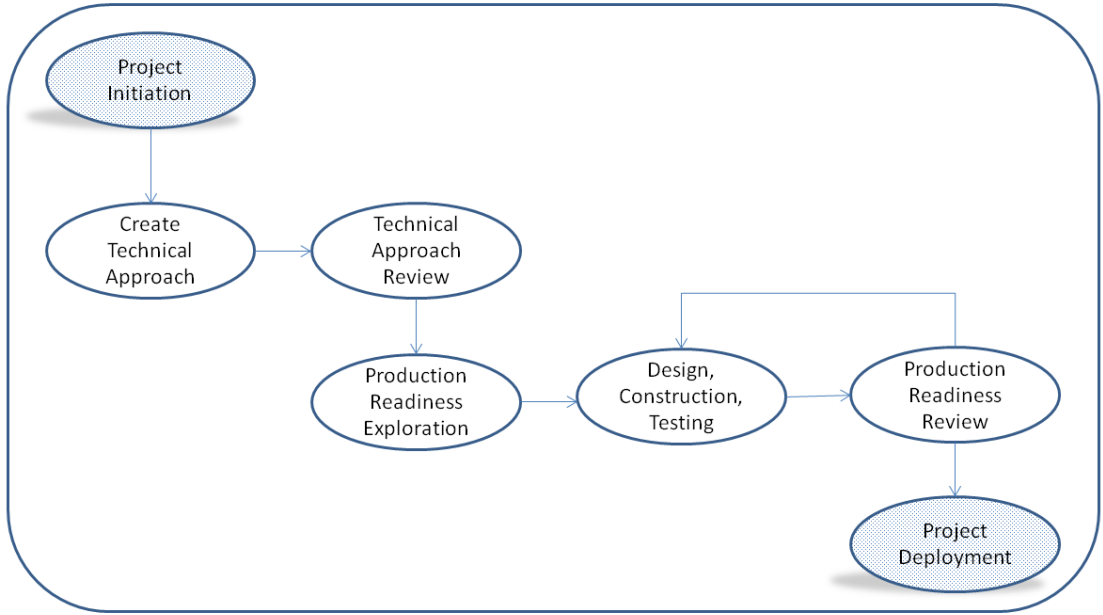

*Figure 1 – High Level Process Flow for Reviews*

## Technical Approach Description

The Technical Approach Document is produced early in the lifecycle in the Analysis Phase. The document provides information to project stakeholders that allows them to develop an understanding of the proposed solution to the problems identified in the project requirements. This provides a basis from which they can either concur with - or bring opposition to - the proposed solution. It is a tool used to facilitate a discussion regarding the feasibility and appropriateness of the technical solution, and - once approved - is used as a guideline for the development effort.

The Technical Approach Document includes descriptions of the scope and functionality stated by the requirements, and identifies the risks, dependencies, and assumptions that are associated with meeting those requirements. It describes the current state of the architecture/work flow. It puts forth a recommended architecture for the project, and demonstrates how that solution is superior to alternatives. It attempts to present practical solutions that fall within technical capabilities, and helps identify existing deficiencies that limit the ability to provide an adequate solution. It also takes into account the Production Readiness Exploration (PRE) checklist described below while determining the proper solution and provides early identification of the system's expected performance, capacity impact, deployment footprint, and interdependencies with existing systems.

# Technical Approach Review

Once the Technical Approach document has been created a review can take place. The Technical Approach Review (TAR) approach was developed starting in 2003 as part of a standardization effort around architecture development. The review was one process that emerged from this effort and its goals include:

1. Provide independent external review of the architecture early in the lifecycle which provides positive improvements to the architecture and avoidance of issues for the project team.
2. Provide expert advice to the project team from knowledgeable and experienced architects and managers.
3. Conduct consistent review coverage from project to project insuring best practices based designs wherever possible.
4. Document the results in a clear manner for follow up and later analysis.

As part of the TAR a PRE is also conducted. The PRE has a very different focus from the TAR which is focused on the architecture. The PRE instead is focused on requirements for environment, capacity preparation, and production readiness attributes and will be explained further below. The main roles in the TAR include:

Review Lead: Each Technical Approach Review is led by a "Review Lead." The role of the Review Lead is to be the central point of contact and coordinator for all review activities. This role is filled by the Project Manager of the project team.

Author: The Author is the person who wrote the Technical Approach and is technically responsible for the contents. The Author is present to address any questions or issues that arise during the review and provides follow-up discussion for those issues that cannot be addressed during the meeting.

Mentors: The review team core membership will be comprised of senior application architects from across the department. The definition of a senior application architect is an IT member who has successfully acted as an architect on at least two projects. This position is voluntary - an individual must apply for the role as mentor. The expectation is that these core members will act as wise, supportive, and influential counselors to all project teams.

SMEs (Subject Matter Experts): SMEs are not permanent team members, but serve as reviewers when requested by the Review Lead. The selection criteria for these temporary reviewers will be based upon their knowledge of the domain, architecture, operations of system under review, and their availability.

In terms of the actual mechanics of the TAR, once it is organized for the level of detail appropriate for the project at hand, review members spend time reviewing the artifacts in advance especially the Technical Approach document. During the review the Project Architect or Author provides a high level overview of the application. Then the Project Architect or Author provides specific details of the architecture, starting by walking through the Technical Approach document. The Review Team raises questions, mostly based on the questions/answers contained in the established Architecture Review checklist, but also ad hoc questions stemming from the content of the presentation and the nature of the application. As a result of this interaction the Technical Lead is often assigned Action Items raised as a result of the review. Minutes are published and it is expected that the issues are resolved and recommendations provided are addressed.

# Production Readiness Exploration

As part of the TAR we also carry out a PRE (Production Readiness Exploration). This review came about in order to prepare solutions for the production environment but also to give support teams an early view into those solutions that were in development. In order to help prepare for new releases coming into production a checkpoint with the development teams was established, called the PRE. The PRE is held during the project's Analysis Phase in conjunction with the Technical Approach Review. During the PRE, the Production Support Team (PS Team) receives a presentation from the development team of the system's expected performance, reviews its expected capacity needs, understands the deployment footprint, and identifies the interdependencies with existing systems.

The goal of the PRE is to alert the PS Team of any major upcoming releases and their ramifications and to allow the PS Team to ask questions around support requirements that the development team needs to consider.

The PRE is a "non-binding" review. This means that the issues uncovered are not expected to be addressed as part of the analysis phase itself. Instead any issues noted need to be addressed during the design and construction phases and will be looked at again in the PRR (Production Readiness Review) which is the final technical review before production launch.

The way the review is held is by utilizing a standard checklist which has been developed over several years of practical use. Aside from logical questions, when there have been problems in production the checklists were updated to cover those types of problems and prevent them on future projects. The development team emails the completed PRE checklist in advance of the review meeting. Ad hoc questions can also be pursued as seems reasonable.

The members of the review team are drawn from the "DevOps" staff covering database and systems specialties as well as production application support.

## Production Readiness Review

Following the Design and Construction phase (or many of them in an Agile model) eventually the project enters final testing. At this time a PRR (Production Readiness Review) is scheduled. Unlike the PRE the PRR is binding. This means that if issues are found in the PRR they need to be addressed in order for the project to advance to a production release. For example, if storage or firewall requirements identified earlier are not in place they need to be met in order for the project to stay on track for a release. These reviews also drive compliance with the updates to our system architecture diagrams. If a project is modifying the computing environment the review process ensure that these artifacts are updated.

The PRR works very much the same way as a PRE. However, the "Production Readiness Review" (PRR) happens during the project's Test Phase. The PRR focuses on the immediate release requirements of the application. This involves a detailed review of the application and covers what might be done in transition planning. It also includes final checks for preparedness of both the Production Support (PS) team and the development team in making the release ready for production. For the review a checklist of questions is published to the development team in advance. The development then submits the completed spreadsheet. For any sizable project a review of this checklist will be held. For small incremental support releases we have an automated workflow using Microsoft Sharepoint that advances the checklist to the review members who vote on whether a meeting is required or the submitted checklist is sufficient. This way we cut down on overhead in conducting reviews for releases that have no change the computing environment except for a software version change. This tool also captures the signoff status of the review.

A request for a PRR review is initiated by the project's Project Manager. Project team members included in the review include the Project Manager, Technical Manager, Project Lead, Lead Developer, and QA Lead. The reviews are comprised of the systems and database representatives as well as application support specialists. Minutes from these reviews are managed through the checklist spreadsheets. Follow through on changes are required of the Project Managers.

## The Checklists

In order to better illustrate the details of the reviews, excerpts from the review checklist templates are presented below. Each of these checklists were developed through real world usage and do not represent an abstract set of good to have questions or criteria. In many cases, questions on these checklists were learned only through failures in the development process resulting in problems reaching production. Based on such experience these checklists have evolved over a number of years of practical use. By sharing them here in part it is hoped they will provide value to others hoping to develop or calibrate their review practices. The first group represents a sample from the Technical Approach Review:

1. General Architecture Considerations
    - Is the design feasible from a technology, cost and schedule standpoint?
    - Have known design risks been identified, analyzed, planned for and mitigated?
    - Have all appropriate artifacts been generated?
    - Does the level of formality match the project size, goals and team expertise?
    - Have any risks with the design been refined based on prototyping or implementation feedback?
    - If possible, are past, successful designs re-used?
    - Are the proposed programming language(s) and technologies approved and are they consistent with our environment?
    - Is the architecture (including data flows, logical flows, states, interfaces, etc.) clearly described and documented?
    - Does architecture take into account current system or environment constraints (that cannot be avoided)?
    - Are error and exception handling consistent throughout the architecture and, if appropriate, the targeted environment?
    - Is the entire system environment described, including hosting HW, 3rd Party software and all internal and external interfaces?
    - Are all build vs. buy decisions included and justified?
    - Have all major algorithms been documented and justified?
    - Are all performance attributes, assumptions and constraints clearly identified?

2. High-Level Design Considerations
    - Does the design have conceptual integrity?
    - Can the design be implemented within our technology and environment constraints?
    - Does the design used standard techniques and avoid exotic, hard-to-understand elements?
    - Is the design extensible? (i.e., will it support future changes or expected changes)?
    - Does the design use existing, shared or shareable, components?
    - Is the use of shared resources indicated?
    - Are component boundaries well-defined, including functionality and interfaces?
    - Are the components loosely coupled or tightly coupled?
    - Does the design handle every possible state or behavior?
    - Are all major data structures described and justified?
    - Are major data structures hidden by an abstraction layer?
    - Are all inputs to methods/interfaces necessary and sufficient?
    - Are all outputs used by a component or method necessary?
    - Is there a strategy for handling abnormal termination?
    - Is there a uniform strategy for handling errors?
    - Does the design leverage best practices such as design patterns'?
    - Does the design separate user-interface, data and business logic from each other?
    - Does the design encapsulate the underlying data structure, from the UI and the business logic?

The next group represents a sample from the PRE:

3. Systems Engineering Readiness
   - Do the new applications need SSL or not?
   - Do the new applications require Load Balancing?
   - Do the new applications need new DNS entries?
   - Are there any Firewall request?
   - Are there any NAS requirements?
   - Are there any Batch requirement?  (run book should be filled out)
   - If batch, at what time of the day will it be scheduled to run?
   - If batch, who will be required to sign off the changes to the run book?

4. DBA and SE Readiness
   - Which Database instance does application use?
   - If new Database instance, which version is the target?
   - Which OS hosting platform is used by DB?
   - Which Application User IDs will be used (new or existing)?
   - Are any Database links required?, if so, specify
   - Which new database features are required by the project?
   - What are the application touch-points with other systems?
   - What are the DB touch-points with other systems?
   - Have Data Volume estimates (monthly, yearly growth) been created?
   - Have user volume estimates been created?
   - Have Archiving and Purging requirements (daily, monthly, yearly) been accounted for?
   - Has application pool, DB connection pool, requirements been considered?
   - Has CRUD matrix (per table, frequency and I/O behavior) been created?
   - Has DB optimal performance been considered (e.g., De-normalization (Pros and Cons explained ), Indexes, Partitioning)
   - Are there any DBMS jobs? If so, what is their frequency?
   - Has SQL and SP code review been scheduled?
   - Who is required to execute the rollback?
   - Has the roll-back plan been tested?

Finally, the last group comes from the PRR checklists:

5. Pre-Conditions to Production Readiness Review (PRR)
   - Are Business Requirements signed-off and up-to-date?
   - Are Functional Requirements signed-off and up-to-date?
   - Is Architecture Documentation signed-off and up-to-date?
   - Is Build Documentation up-to-date?
   - Is there an automated build for application?
   - Has all new or modified code been reviewed and signed-off?
   - Have all touch-points been identified? (Batch, 3rd Party, external system, etc.)
   - Will all touch-points be tested in QA? (Batch, 3rd Party, external system, etc.)

6. Capacity Planning Readiness
   - Have Load Capacity/Load Requirements been signed off?
   - Has host platform performance (ex: CPU, Memory, .Net process, Build and migration process, etc.) been estimated?
   - Do Load Test results show: max transactions per second, step level and transaction breakpoint metrics?
   - Has Storage Capacity Analysis been done?

7. Performance Readiness
   - Had all CRUD functionality been load tested?
   - Using Instrumentation data: Has all SQL performance been analyzed? (and their execution plan)
   - Has execution times of each new/modified SQL/Stored Procedure been reviewed and signed off by DBA?

8. Batch Applications
   - Have any new Batch applications been created as part of this release?
   - If so, has the performance been tested or estimated?
   - If so, has the new batch been tested in an environment in which all current batches are run at the same time?

9. Deployment Readiness
   - Is new SSL required?
   - Is new Load Balancing required?
   - Will any additional data files be stored in NAS?
   - Are new DNS entries required?
   - Are any new Firewall requests required?
   - Is there an application and DB rollback plan?

10. Post Go-Live Maintenance Readiness
    - Have DB Backup and Recovery Requirements been documented, reviewed and signed off?
    - Have DB Statistics and DB Reorg frequencies been documented, reviewed and signed-off?
    - Are there any purging/archiving requirements?
    - Are any IP Addresses being changed?
    - Are database instances being added?
    - Are we Adding/Modifying the clustering or load-balancing of any servers?

## The Results

The Technical Approach Review was put in place several years ago. Since then all major applications that were developed with a Technical Approach document have held a TAR. This equates to dozens of reviews over the years. As part of these reviews numerous issue have been identified and corrected early in the project lifecycle. On several occasions due to the cross team expertise brought to bear on the problems very creative solutions were contributed in these meetings. One such case was where the architect was going to custom build a data bridging capability but one of the reviewers pointed out this was a standard functionality in one of our COTS tools available in the environment. At other times the focus around computing performance in these reviews has aided projects in planning for better designs.

During the first half of 2012 there have been 6 such reviews. During these particular TARs there have been few technical issues reported. They have, however, served the purpose of informing the technical organization of the design plans coming down the pipe. However, during the PREs and PRRs there have been several issues raised with the project teams. One of the major items identified during a PRE was the lack of solid requirements from a third party partner. This was flagged as a key concern for the project.

As for the PRRs, there were numerous issue identified. First of all, during this time period there were 55 PRRs as all support releases require a PRR (but will not require a TAR) and there are many more support releases than project releases. The majority of these reviews were done "offline" meaning that as defined by an automated workflow the domain experts review the release content against the PRR checklist and if it passes muster no review meeting is needed. However, in the first half of 2012 there were a total of 10 releases that did require a review meeting as determined by the domain experts. In these cases there were a variety of issues discovered that needed some response by the release team. In some cases database SQLs had not been reviewed, in another case changes to the platform needed to be reflected in our design reference diagrams, in another case rollback scripts that were needed had not been created, and finally there was a case where the batch deployment and operations approach needed to be reworked. Naturally the PRR cannot catch every mistake but it does tend to eliminate the common problems found in the checklist scope above.

## Discussion

Any process can be improved. The review process we have in place is not an exception to this. The strengths of the process are that it is well known and understood by all the participants (although when we bring new subcontractors on board we need to train them on what is required). Also, the process is debugged. We have an automated means of scheduling reviews and many reviews are conducted "offline" so that meeting time is reduced. Finally, the checklists (as mentioned) were developed over years of practical use. This makes them relevant and to the point. There is little argument about the nature or content of the reviews (this was not true at the beginning when we had to help people get used to them and see their value).

Where there is room for improvement is in the preparation time. For the reviewers it does take time to come up to speed on the designs especially in the TAR. The project teams and architects have been working on the issues to be reviewed typically for weeks or months and are very well versed in the problem space and the solution being proposed. It is sometimes frustrating for them to take the time to educate the reviews with the background and the details required for them to make a useful contribution. This is especially so if the reviews are scheduled for an hour. This makes it nearly impossible for the reviewers to come up to speed adequately so a balance must be struck between the pre-work and the duration of the review. One way of managing this is to encourage the project teams to review the checklists in advance and prepare answers so that it is easier for the reviewers to look for any anomalies.

## Conclusions

It has been nearly 20 years since my first exposure to an architecture review. Since then I have participated in dozens of reviews covering process, architecture, and operations and led many as well. I have also watched as these methods have become standardized and routine in many parts of the industry. In this paper what I have tried to do is layout in some detail the approach that we use at Wolters Kuwer Corporate Legal Services. This approach mimics the typical architecture reviews found in the industry but provides them in a light weight model. The approach also introduces the concept of a Production Readiness Exploration to get people thinking about the realities of launching a system or application into production early in the lifecycle. This is followed by a Production Readiness Review which is used as a gating process to ensure true application readiness for production. These approaches were presented in some detail and with corresponding checklists developed in real-world conditions. It is expected that others can emulate this approach or improve on it but certainly find value in the unique combination of reviews, checklists, and process flow.

## Acknowledgements

I owe a large debt of gratitude to those people who ran and taught architecture reviews at AT&T Bell Labs. The list is very long but of special note is the late Ted Kowalski who was a leader in this area. I also appreciate all the people at Wolters Kluwer who have cooperated to create this review process and document carefully the gotchas that can impact a project. This is another long list which includes Kevin Lane, Louisa Tam, Manolo Alvarado, Gary Ma, and many others. Finally, John Kostecki who administers many of these reviews assisted in the perpetration of this paper by providing some of the source material used.